# Manual Calibration System for Daya Bay Reactor Neutrino Experiment

H. X. Huang*, X. C. Ruan, J. Ren, C. J. Fan, Y. N. Chen, Y. L. Lv,

Z. H. Wang, Z. Y. Zhou, L. Hou, B. Xin

China Institute of Atomic Energy, Beijing, China

C. J Yu, J. W. Zhang, Y. H. Zhang, J. Z. Bai, H. L. Zhuang，W. He

Institute of High Energy Physics, Beijing, China

J. L. Liu

Shanghai JiaoTong University, Shanghai, China

E. Worcester, H. Themann

Brookhaven National Laboratory, USA

J. Cherwinka, D. M. Webber

University of Wisconsin, USA

**Abstract**: The Daya Bay Reactor Neutrino Experiment has measured the last unknown neutrino mixing angle, $\theta_{13}$, to be non-zero at the 7.7$\sigma$ level. This is the most precise measurement to $\theta_{13}$ to date. [1,2]. To further enhance the understanding of the response of the antineutrino detectors (ADs), a detailed calibration of an AD with the Manual Calibration System (MCS) was undertaken during the summer 2012 shutdown. The MCS is capable of placing a radioactive source with a positional accuracy of 25 mm in R direction, 12 mm in Z axis and 0.5° in $\Phi$ direction. A detailed description of the MCS is presented followed by a summary of its performance in the AD calibration run.

## 1. Introduction to the experiment and an Antineutrino Detector(AD)

In the Daya Bay (DYB) reactor neutrino experiment, the energy and rate of anti-neutrinos from six 2.9 $GW_{thermal}$ reactors were measured with eight antineutrino detectors (ADs) deployed in two near (flux-weighted baseline 470 m and 576 m) and one far (1648 m) underground experimental halls[1,2].

We detect anti-neutrinos using the Inverse Beta Decay (IBD) reaction

$$\overline{\nu_e} + p = e^+ + n. \qquad (1)$$

Referring to Fig. 1, each AD has a 3-zone design in a nested "Russian doll" scheme. The inner-most region is a 3 m diameter by 3 m tall inner acrylic vessel (IAV), filled with 20 tons of liquid scintillator (LS). This LS has an added 0.1% of Gadolinium to create a doped liquid scintillator (Gd-LS). This is the target zone and is designed to eliminate the need for a fiducial cut of the data [3]. Surrounding the target zone is 20 tons un-doped LS, held by a 4 m diameter 4 m tall outer acrylic vessel (OAV). This zone functions as a gamma catcher of the photons from IBD events that occur near to the boundary of the target zone. Outside of the gamma catcher is the main vessel, a 5 m diameter 5 m tall stainless steel vessel (SSV) with 192 photomultiplier tubes (PMTs) and reflectors at top and bottom as well as other supporting structures. This outer volume is then filled with mineral oil which absorbs radiation from decays in the glass of the PMTs, the SSV, as well as the surroundings of the AD.

*Corresponding author. Tel: +86 01 69357116, Email address: huanghx@ciae.ac.cn

On the top of each AD, there are 3 overflow tanks (OVT) to accommodate any expansions or contractions of corresponding liquids due to temperature changes and potential deformations during transportation. Also on top of the ADs are three automated calibration units (ACUs) [4]. The ACUs and OVT are key elements to guarantee the precision of our measurement. The ADs are set in water pools to decrease the radiation background and to provide a muon veto. More about the design and performance of the ADs can be seen in Ref. [5].

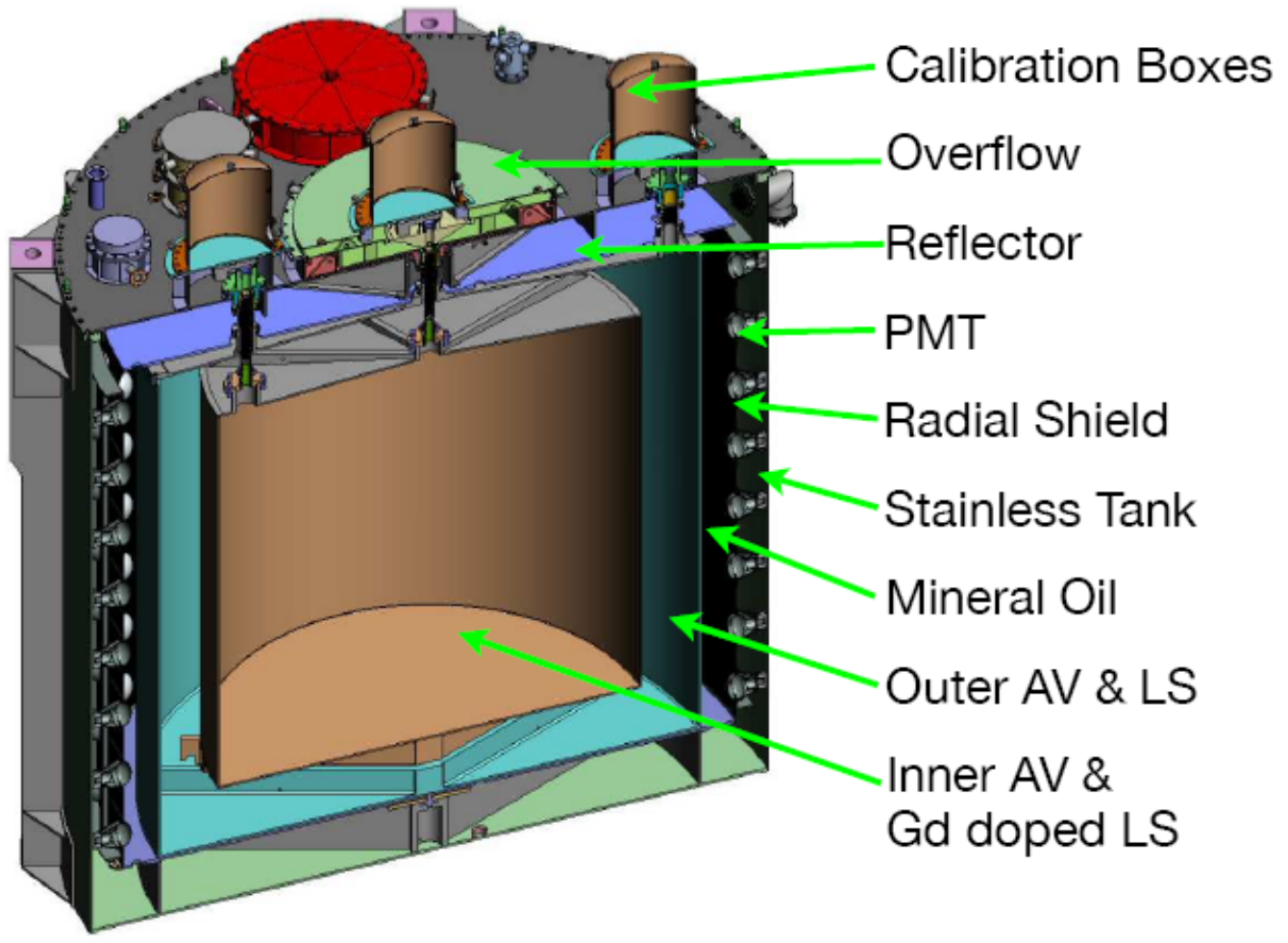

Fig. 1 The structure of the DYB AD

## 2. Overview of the MCS

The ACUs are placed at three radial positions: the center of the target zone (IAV), the outer edge of the target zone (IAV) and the gamma catcher zone (OAV). The sole degree of freedom for the ACUs is along the Z axis. The simplicity of the design allows the system to be automatic and local to the detector for frequent calibrations. The compromise of this design is that it cannot fully explore variations in detector response over the full volume.

In the AD, the deposited energy from particle interaction is reconstructed based on the amount of scintillation lights (total number of photoelectrons measured by the PMTs). The conversion is known to be nonlinear due to the so-called scintillation quenching and Cerenkov effects [6]. Radioactive sources with different characteristic energies are natural choices to calibrate this non-linear dependence. Variations in the response are expected due to differences in optical absorption of the liquids, variations in reflectance of the reflectors and differences in distance and performance of the PMTs. These variations were called non-uniformity.

The motivation of the MCS was to explicitly measure the signal response to AD as a function of position throughout the target zone. The AD response included but not limited to non-linearity, non-uniformity, energy resolution, detection efficiency and so on. In addition, the detection of neutron captures is a key aspect of an AD. There are several aspects to this including the capture time, the ratio of captures by Hydrogen vs. Gd and the behavior of captures at the boundary of the target zone.

The design requirements were defined as:

1) The calibration sources should be deployed to the fiducial volume of the IAV through the 50 mm diameter central calibration port.

2) According to the DYB project technical design [7], the precision of 1 MeV peak should be smaller than 1% and 6 MeV peak should be smaller than 2%. For this, the calibration sources should be accurately deployed with a position accuracy of 20 mm so as to achieve the peak precision of 0.5% [8,9].

3) The position of the source can be well controlled, determined and repeated.

4) All materials used must be compatible with Gd-LS (Inside the IAV) or pure water (In the water pool) to avoid any possible contamination.

5) Radon gas in the air, $^{222}$Rn in particular, can contaminate the liquid scintillator by introducing long lived decay daughter $^{210}$Pb (22 years half live). $^{210}$Pb's daughter, $^{210}$Po will α decay, which could make neutron background via (α, n) on $^{13}$C in the scintillator. To set the scale, 100 Bq$^{210}$Po decays would introduce a 0.1/day/AD background (0.1% of the IBD signal rate in the far detector)[7]. Therefore when the AD is opened for the MCS deployment, both AD and the MCS needs to be flushed with dry N2 to avoid direct contact with air. The MCS needs to be sealed with the AD during the deployment.

6) Though called "manual", the deployment and movement of sources in the AD should be automated to a large extent to have better control and avoid human intervention.

7) There should be no chance of damage to AD, in particular the source must not become trapped inside the AD.

8) The signature of a neutrino interaction in the DYB ADs is a prompt positron with a minimum energy of 1.022 MeV plus a delayed neutron. About 90% of the neutrons are captured on Gadolinium, giving rise to an 8 MeV gamma cascade. The radioactive sources used should provide an energy range covered 1-8 MeV so as to delineate the AD response.

9) To provide information to the neutron captures as a function of position, a neutron source should be used in the MCS.

10) The source rod should have minimum shadowing or other additional effect on the normal operation of the detector and the control system should not affect the PMT signals.

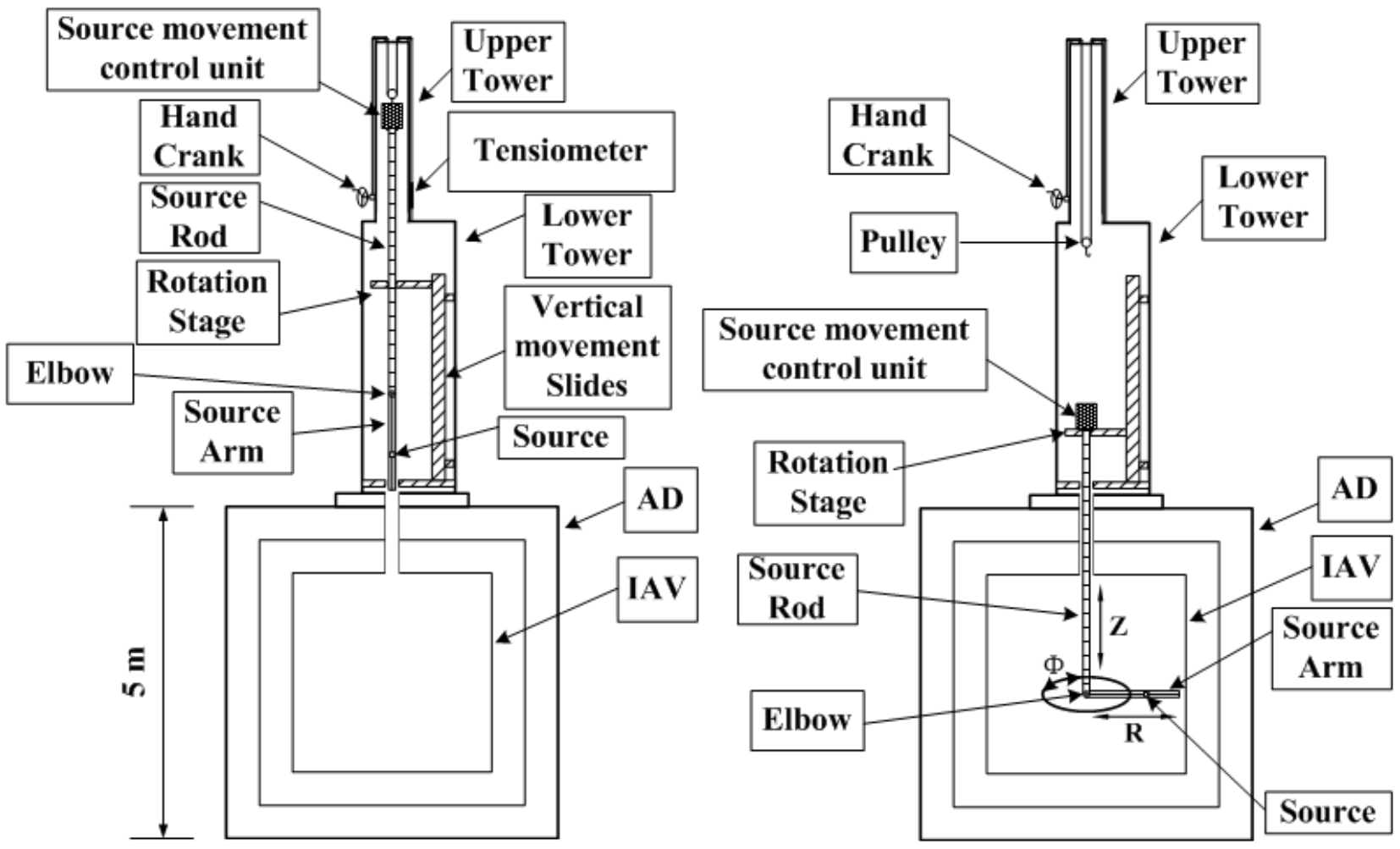

(a) (b)

Fig. 2 Schematic of MCS

(a) Scenario before the source rod insertion into AD. (b) Scenario after the insertion.

A schematic of the MCS installed on an AD is seen in Fig. 2. The MCS was composed of three main components: calibration tower, Source Rod Assembly (SRA, Fig. 3) and control system. The calibration tower is divided into upper and lower towers. The upper tower allows the SRA to be completely retracted into the tower assembly when transporting the MCS. The lower tower contains the mechanism that controls the vertical and rotational motion of the SRA.

In this design, an articulated source arm was used. The source rod and the articulation point, the elbow, were made of stainless steel (SS) and coated with Teflon to prevent Gd-LS from directly contacting SS. The source arm was made of an Acrylic tube. The source arm could be opened 90° relative to the source rod and also could be retracted 180°. The calibration source was located inside the source arm and can be moved along the source arm. In this way, the source could be deployed to almost any position of the IAV except the edge very near the wall of the IAV.

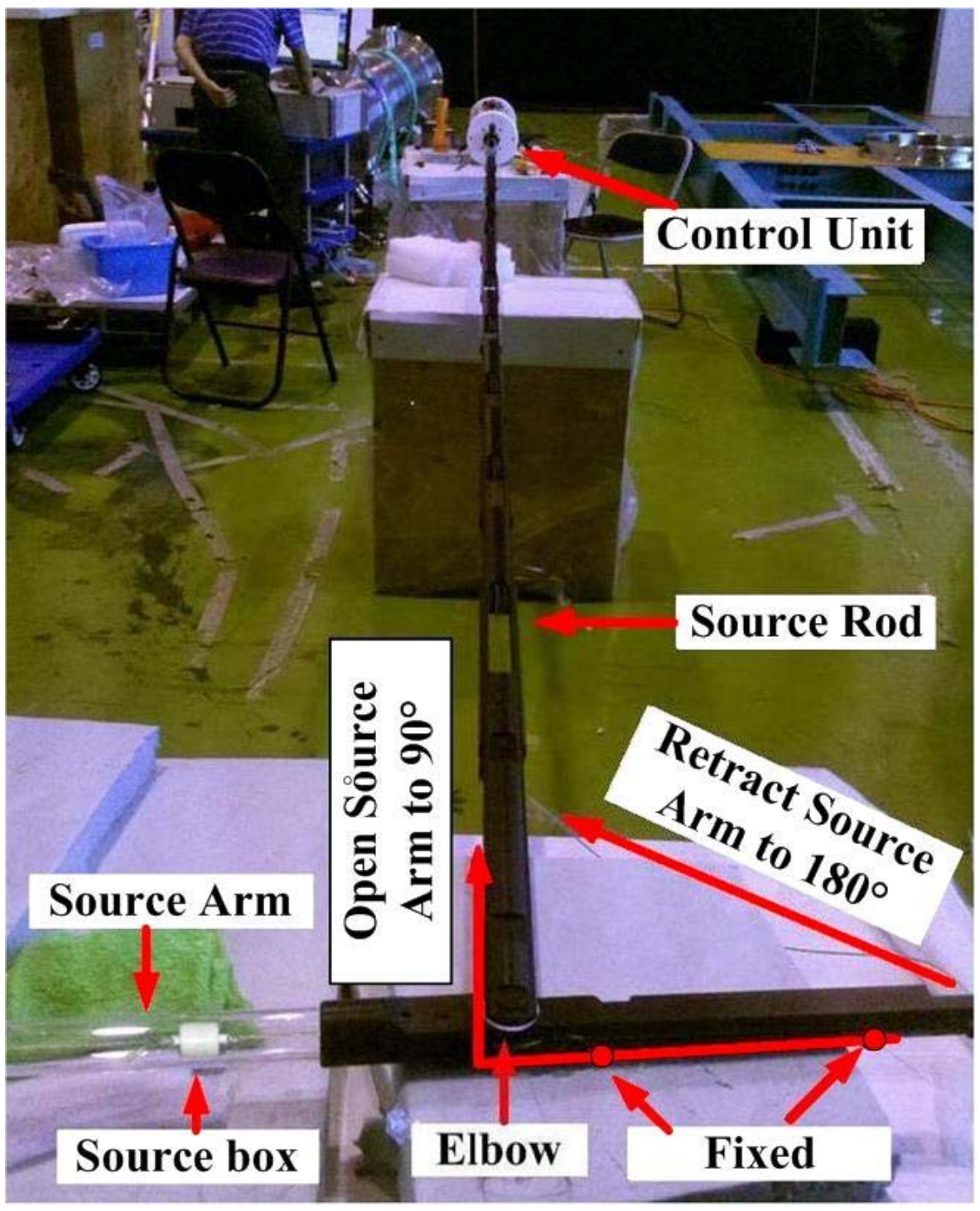

Fig. 3 Component of SRA

The working principle of the MCS is described as follows:

1) All components were cleaned before assembly and installation. The MCS was assembled in the clean room and then transported to the experimental hall for installation.
2) The MCS was purged with nitrogen gas after assembly, and continuously flushed by nitrogen gas after its installation onto the AD. This prevented air, Radon in particular, from going into the AD during calibration. Oxygen would degrade the quality of the liquid scintillator [10].
3) A leak check was performed before and after the MCS installation on the AD to ensure there was no water leakage into the AD.
4) Prior to MCS installation the water pool was drained to expose the top of the AD. An access platform was installed onto the AD and the MCS lowered through the platform (Fig. 4). Once the MCS was installed onto the AD, the water pool was re-filled bringing water up to 2 m above the top of the AD.

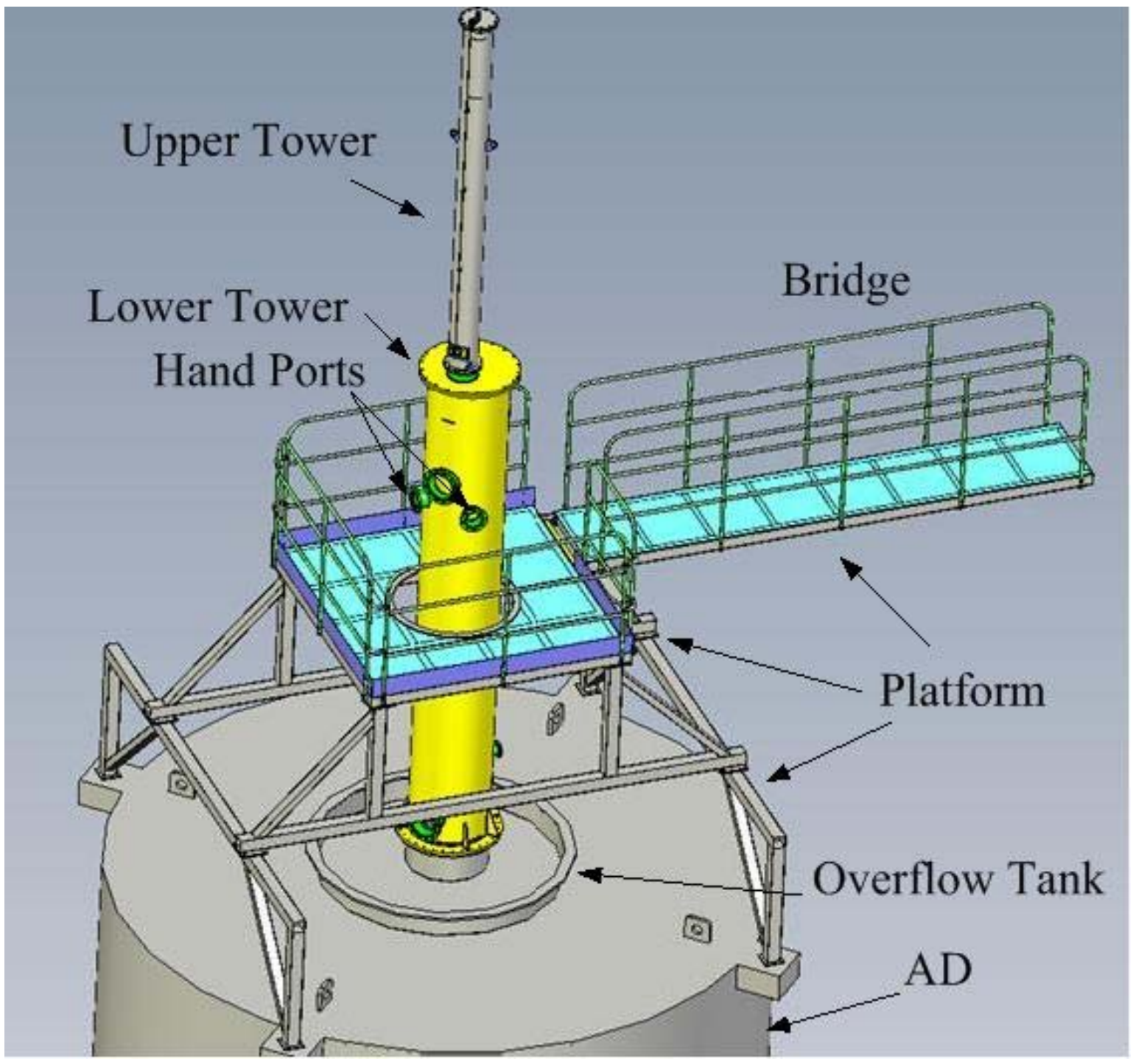

Fig. 4 3D model of the MCS

5) Upon completion of the pool filling, the SRA was lowered into the lower tower. Referring to Fig. 2(a), the assembly is suspended by a SS wire rope while in the upper tower. The assembly was lowered using a hand operated winch, the tension meter allowed the operator to detect obstructions. The SRA was lowered until the control unit came to rest in a keyed slot of the rotation stage. The SS rope was disconnected and retracted. At this point the source arm and elbow had penetrated the IAV via the central calibration port.
6) While the rotation stage was in the upper position as in Fig. 2(a), the control system was connected to the control unit and all articulations were verified. After lowering, the source arm was opened to its 90° position by hand cranks on the control unit. All movement of the assembly was then done by the control system via a PC.
7) With the source arm open to the 90° position, as shown in Fig. 2(b), the rotation stage supporting the control unit can move up and down (Z axis) controlled by a servo motor. The source arm can rotate around the Z axis driven by a stepping motor (Φ direction). The source was sealed with a Teflon container and was able to be moved from one end of the source arm to the other end driven by a motor through a rope system (R direction). The motor was installed inside the control unit.
8) When the calibration was finished, the source arm was retracted from 90° to 180° and the SRA was lifted into upper tower.

## 3. Calibration sources

The MCS used a combined source including of $^{60}$Co and $^{238}$Pu-$^{13}$C (Fig. 5). High activity reduces data acquisition time but this must be balanced by the need to avoid saturation effects of the PMTs and DAQ system. The radioactivity of the $^{60}$Co source was 150 Bq and the $^{238}$Pu-$^{13}$C was about 2000 Bq for neutrons and 40 Bq for 6.13 MeV gamma rays respectively. These sources covered the energy range from 1 MeV to 8 MeV.

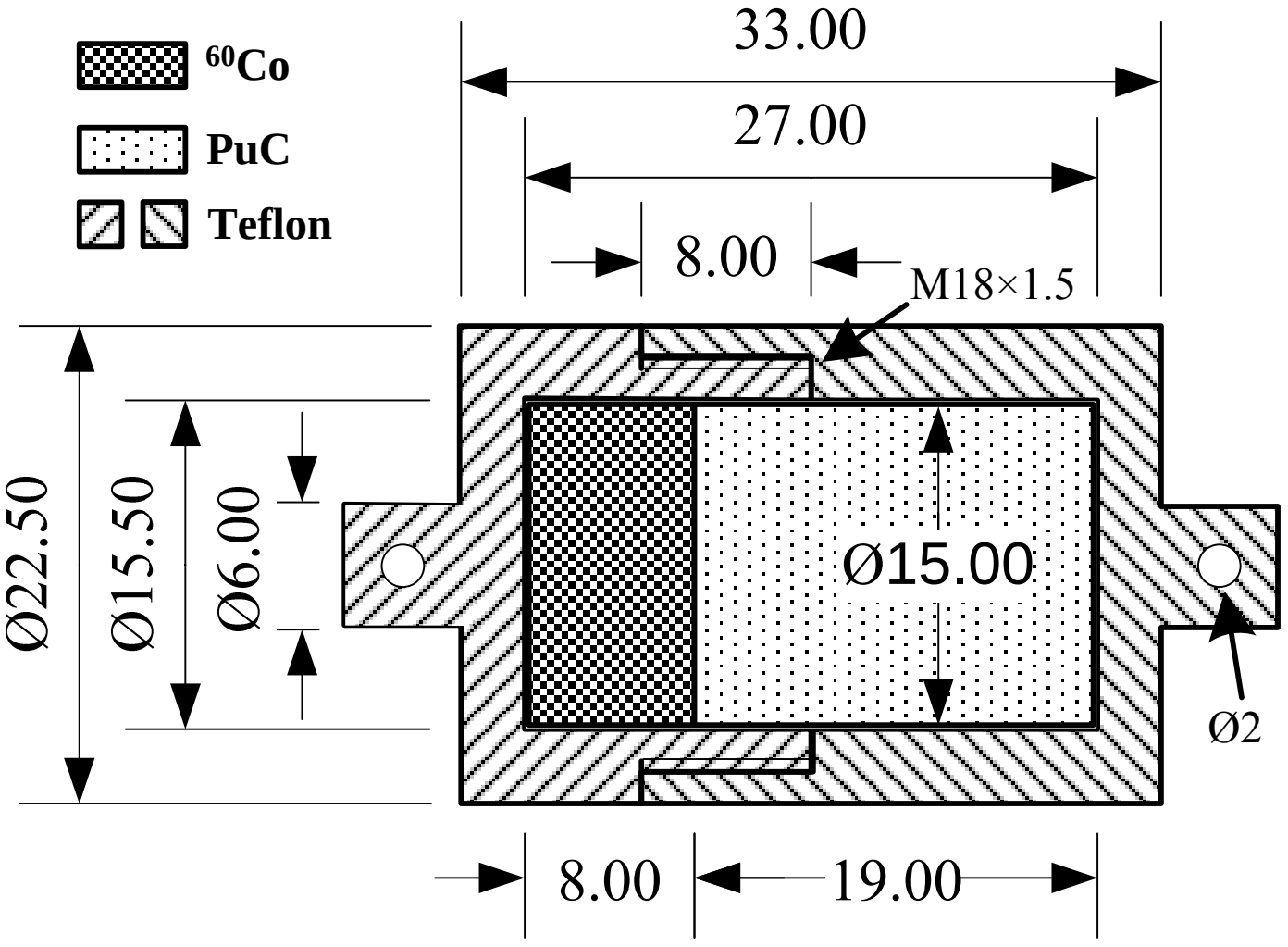

Fig. 5 Geometry of the source enclosure

There are two key energy points in the DYB neutrino experiment: one is 1.022 MeV, which is the lowest energy of the prompt positron signal from the neutrino events; the other is 6 MeV, which is the minimum energy selection cut of delayed neutron signal candidates from the neutrino events (Fig. 14). More information about the neutron spectrum of the $^{238}$Pu-$^{13}$C source can be seen in Ref. [11].

The neutrons from a $^{238}$Pu-$^{13}$C source can also simulate the neutron capture signals in the AD. In addition, the $^{238}$Pu-$^{13}$C source can also provide five other energy points in the energy region of 1 - 8 MeV： 1.02 MeV from an electron pair effect, 2.2 MeV from gamma ray emission of the H(n, γ) reaction; 4.43 MeV from the $^{12}$C(n,n')$^{12}$C* reaction; 4.95 MeV from gamma ray emission of the $^{12}$C(n, γ) reaction, 6.13 MeV for $^{16}$O* decay and ~8 MeV from the Gd(n, γ) reaction (Fig. 14). The combined source covers the energy range from 1-8 MeV, so it is suitable for the study of the AD response and other properties of the ADs. The signal rate for an energy threshold of 0.4 MeV was measured to be about 3 kHz. Most of the events are delayed coincidence pairs in which neutron capture serve as the delayed signals.

The $^{60}$Co source served as a cross check to compare MCS with ACU.

## 4. The details of the MCS components

### 4.1 The SRA

The SRA (Fig. 3) is a critical component of the MCS. It was used to deploy the source and it is the only part that directly contact with Gd-LS. The design of SRA has many constrains:

1) The diameter of source rod and source arm must be smaller than the central calibration bellows ($\phi$= 50 mm) to allow the insertion of the assembly into the IAV.

2) The stiffness of the material must be sufficient to resist deflection and deformation of the SRA which would directly contribute to position accuracy.

3) The source arm must be transparent to the source so as to minimize shadowing.

4) The SRA must be of minimum mass to reduce the load on the servo motors and minimize the damage risk of damage to the IAV.

5) The Gd-LS cannot be "topped up", removal during retraction of the assembly must be minimized.

The SRA is 6.5 m long. To support its own weight and to avoid shape distortion during movement, SS is selected as the material. The surface of the SS is coated with Teflon to be compatible with the liquid scintillator. A series of interlaced slots were cut into the SS tube to reduce its weight (Fig. 3). The outer diameter (OD) of the source rod is 38 mm and inner diameter (ID) is 30 mm. The length of the source rod is about 4.5 m.

An Acrylic tube with 30 mm OD and 24 mm ID was used as the source arm which was compatible with Gd-LS, low mass and transparent to light and the source radiation.

The elbow was used to connect the source rod and the source arm. The angular range of elbow movement is from 90° to 180°. A block of SS on the other end of source arm was used to counterbalance the source arm and reduce the force of opening the source arm. According to a rough measurement in air, the force for opening the source arm to 90° is 13 kg and the force for retracting the source arm to 180° is 3kg. These forces are smaller in the LS due to the positive buoyancy of the source arm.

The source rod was connected to the Source Control Unit (SCU) which in turn rested on the rotation stage in a keyed slot. The control unit opened and closed the source rod and it translated the source along the source arm. In Fig. 6(a), a CAD visualization of the SCU is presented. The rope for driving the source movement along source arm was attached to a grooved wheel. The rotation of the grooved wheel in both directions was driven by a stepping motor through a timing belt so the position is known. A potentiometer was used to provide a redundant readout of the R position of the source. In order to prevent the source movement from going over range, two mechanical limit switches were used (Fig. 6(b)). The motor was powered off when the limit switch was activated. This was done to prevent breakage of the SS rope or the Acrylic source arm. The ϕ0.46 mm SS rope for driving the source movement is sealed with Teflon sleeve (ϕ=0.66 mm with the sleeve). The size of the SS rope was chosen such that it could support the weight of the source arm and source in the event of the source arm breaking.

Source arm opening and retracting were also realized by a rope system. The ropes were attached to the two ratchets installed on the SCU. This operation was done manually through hand ports on the calibration tower. The size of the SS rope used for source arm opening and closing was ϕ1.19 mm, also sealed with Teflon sleeve (ϕ=1.59 mm with the sleeve).

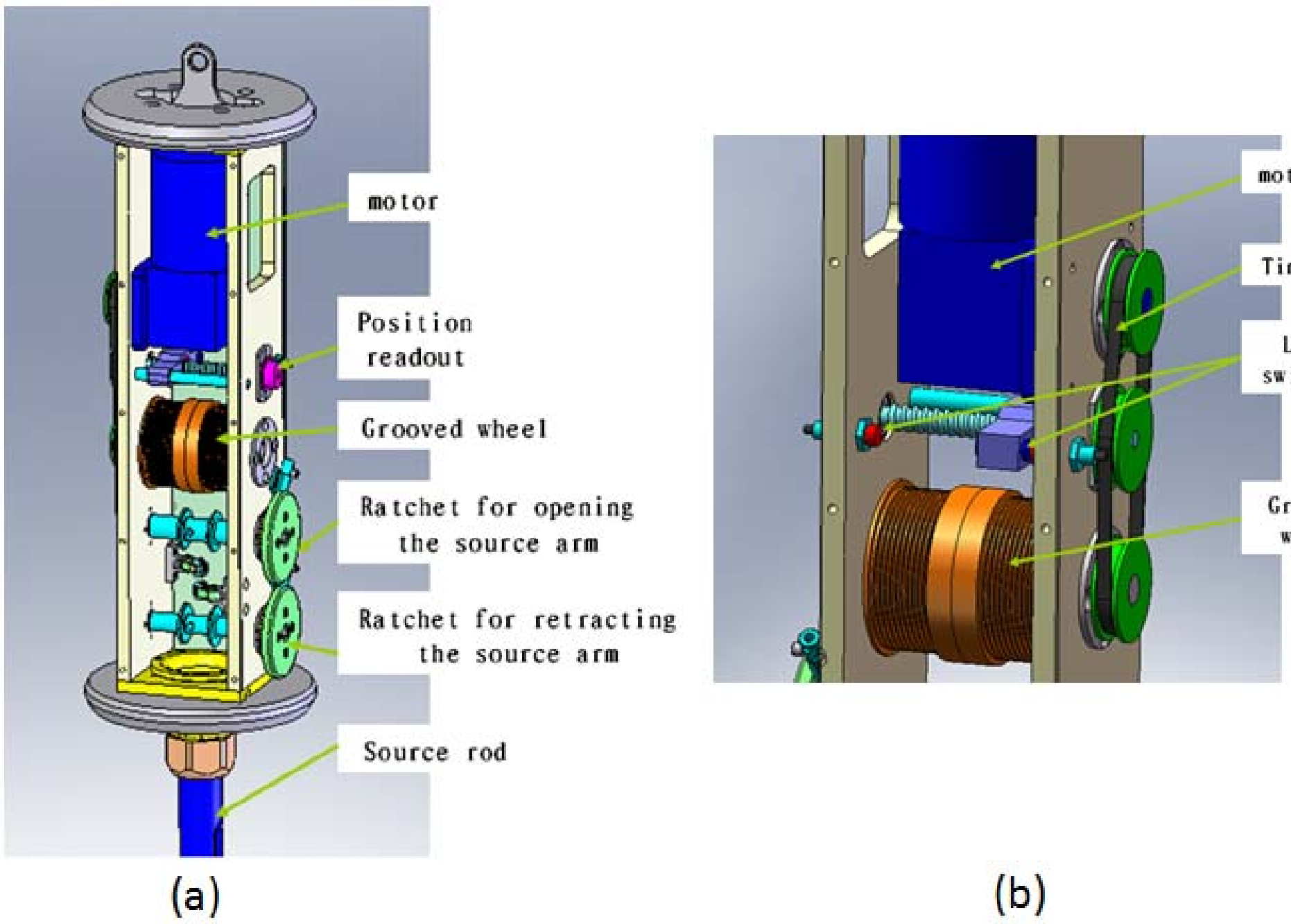

Fig. 6 SCU and the limit switches for source movement along R

### 4.2 The calibration tower

The calibration tower was used to seal and install the SRA, provide the interface to the AD and provide the interface for manual operations. The design requirements for the calibration tower included: 1) Waterproof for the parts under the water; 2) Be compatible with pure water; 3) Light shield.

According to Monte-Carlo simulation [8], one meter height of water above of the AD lid could reduce most of the background from rock and air. 1.5 m of water above the OVT was adopted during the calibration to enhance the signal to noise ratio. A double O-ring seal was used for the bottom flange of lower tower to ensure no water could leak into the AD and to simplify leak checking. The tower assembly passed a leakage check before its installation to ensure no leakages on the flanges. The tower was made of SS for its compatibility with pure water. A light shield was also needed to avoid light leakage into the AD.

The upper tower was used to seal and support the SRA during installation. The total length of the SRA is about 6.5 m. The height of the lower tower is 4.2 m. Therefore, additional upper tower was needed to seal the SRA. A rope system to hang the SRA was also installed on the upper tower. The source rod insertion before calibration and retraction after calibration were realized by rolling the hand crank on the upper tower. There was a spring scale tension meter on the rope system to monitor the tension on the rope during lifting or lowering the SRA. The tension meter was also used as a cushion to smooth the acceleration of the movement.

A vertical movement slide was installed in the lower tower (Fig. 7). The SRA was installed on the rotation stage. Then the rotation stage was able to move up and down driven by a servo motor through a lead screw and slide structure. The rotation stage could drive the SRA rotating with it. The rotation motion was controlled by a stepping motor installed on the rotation stage. Krytox was used as a lubricant to replace the common lubricating grease on the lead screw to prevent any possible contamination of the Gd-LS in the AD.

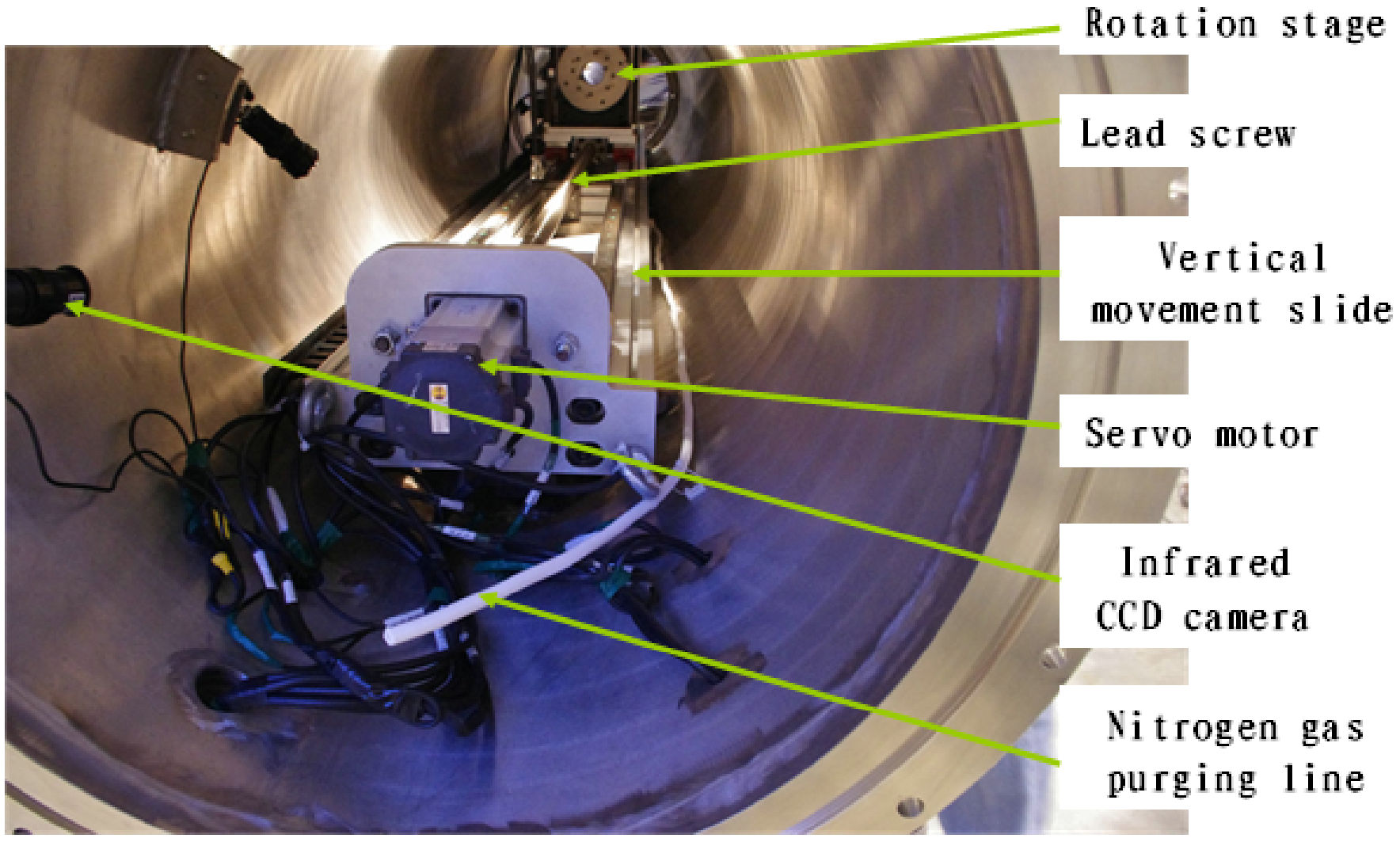

Fig. 7 Inside of lower tower (top view)

### 4.3 Hand ports

As shown in Fig. 4, two hand ports were designed on the lower tower. They were used in the following situations: 1) Installation of the SRA onto the rotation stage during the insertion of source rod into the AD; 2) Connecting and disconnecting the rope for lifting the SRA; 3) Opening and retracting the source arm. The flanges were opened before manual operations and sealed after the operations to keep the nitrogen gas inside and shield the lights from the outside.

### 4.4 Interface to the AD

The calibration tower was installed on the support flange of the OVT. Double O-rings were used between the support flange of OVT and the bottom flange of the MCS. The nitrogen purging line for the AD was connected to the MCS through an ISO63 flange. There were two feedthroughs on the flange, one for gas in and the other for gas out. The operation platform of the MCS was seated on the lifting fixture attachment points of the AD. This can minimize possible interference between the MCS and the AD.

### 4.5 Monitoring system

An optical remote inspection system including three cameras (CCD) inside the MCS (As shown in Fig. 7) and two cameras inside the AD [12] was used to monitor the movement of MCS. The infrared cameras did not affect the PMT signals during calibration.

### 4.6 Operation platform

As shown in Fig. 4, the operation platform was used to help people reach to the MCS for manual operations. Since part of the calibration tower was submerged in the water pool, this platform was necessary. The platform was made of SS for compatibility with the pure water.

### 4.7 Control system

After the control assembly was seated on the rotation stage and the pulley was retracted to upper tower, all motions of the source were performed with the control system.

Three motors were used to control the source movements along three different directions. These motors were controlled by a computer. All motions were protected by limit switches. This was very important to prevent the AD or the MCS from any possible damage. For Z movement, two sets of top and bottom limit switches were used. One set used photo-electrical limit switches and the other set used mechanical limit switches. The photo-electrical limit switches defined the movement range of the source along Z axis. The mechanical limit switches served as an additional protection. This meant that the motion can be stopped safely even if the photo-electrical limit switches failed. The top photo-electrical limit switch also served as home position for the Z axis. It was used to define the Z value of the source. For rotation movement, one photo-electrical limit switch was used to define the home position for the Φ direction. The rotation was protected by limiting the rotation range between 0 and 350 degrees. This could avoid the possible tangling of the electrical cables connecting to the SCU. The source movement along R was also protected by mechanical limit switches and one of the limit switches was served as home position of the R value. In addition to the hardware limit switches, all motions were also protected by software.

A simplified diagram of MCS control system can be seen in Fig. 8. The model of motion control card is a MPC08SP [13] by Leetro Automation Co. Ltd.. The model of the servo motor is MINAS A4 [14] by Panasonic. The model of the stepping motor driver is a DMD403 [15].

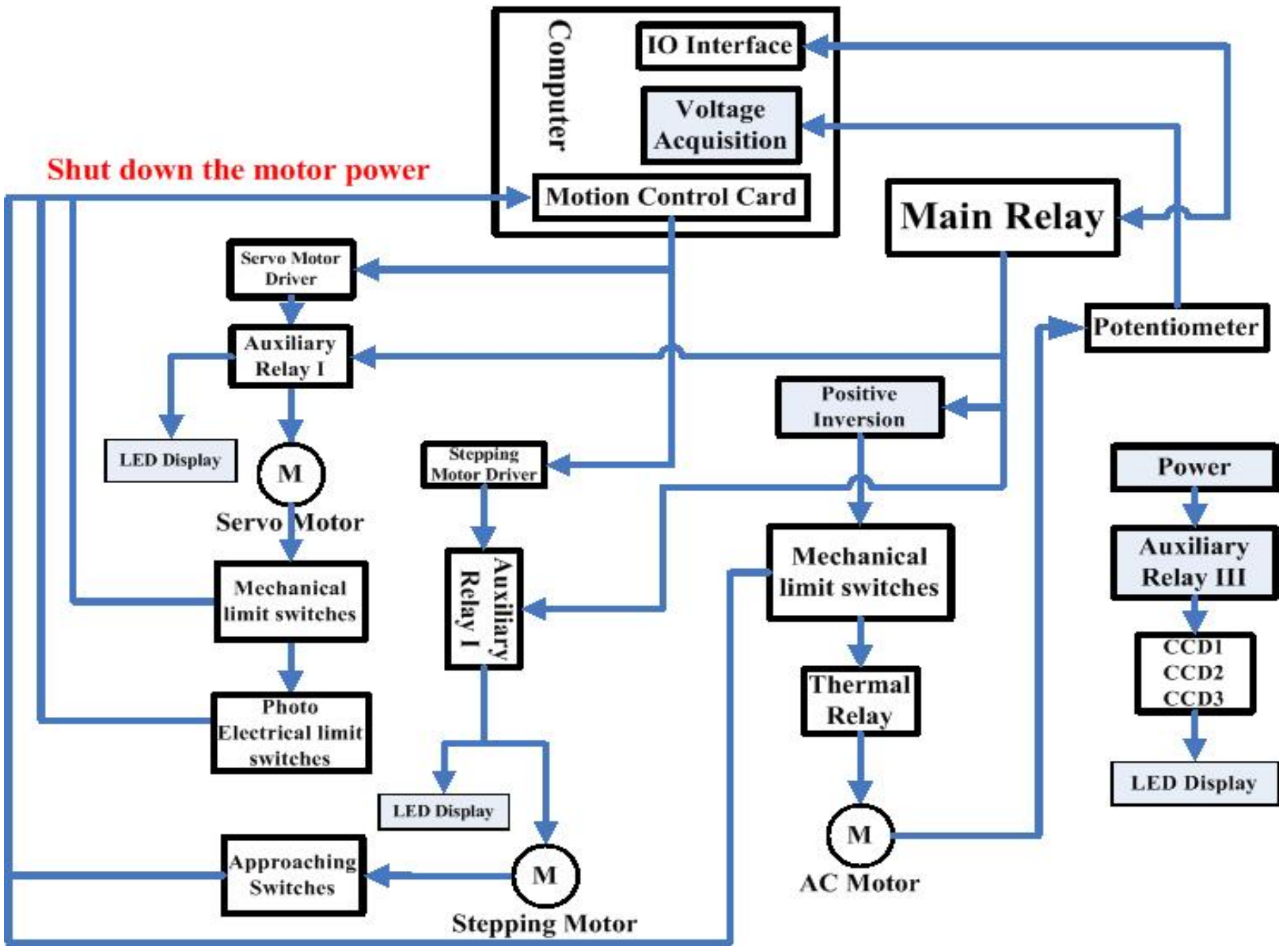

Fig. 8 Logical diagram of MCS control system

Fig. 9 shows the interface of the MCS control software. This software was developed with Borland Delphi 7 which is an object-oriented and visual programming environment [16]. The initial movement parameters such as movement ranges, home positions and potentiometer calibration results can be pre-set under the "Setting" panel. All values of these parameters were checked by the software itself which means that an error message will pop up when the values of inputted parameters were out of range. This could avoid some problems from accidental or inaccurate input. This software can work in two modes: debugging mode and calibration mode. In debugging mode, the source position values (R, Z and Φ) were input manually. In calibration mode, a position list was loaded before starting a new calibration job. And then the source can be deployed to a set of defined positions step by step by clicking the "Continue" button. All motors, limit switches and CCD status can be monitored in this software.

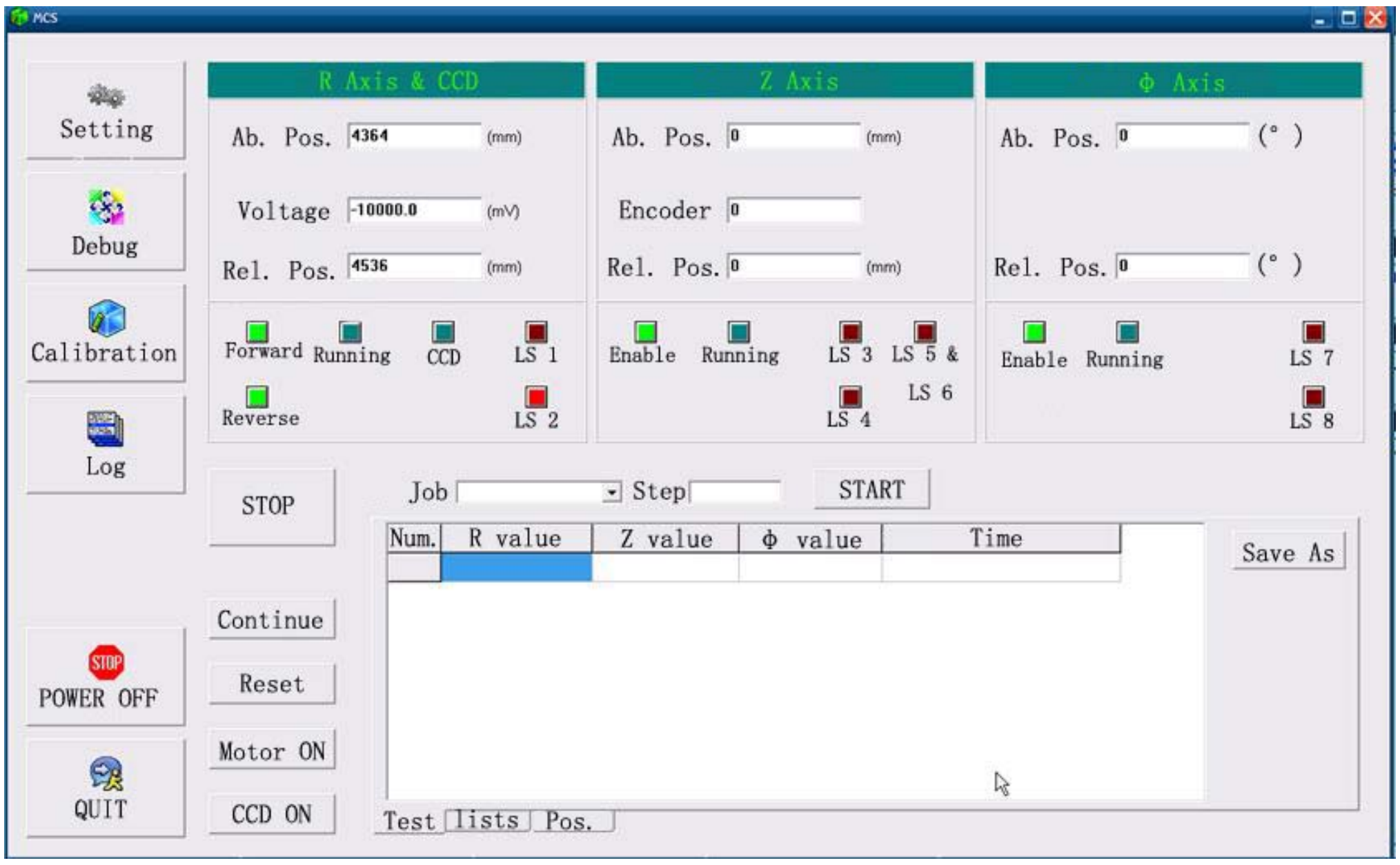

Fig. 9 Interface of MCS control software

All motors were electrically isolated from the AD to avoid introducing additional noise into the PMTs of the AD. All motors were powered off during data taking. The servo motor which was used to control the Z axis movement would automatically self-brake when its power was shut off. This could prevent the control assembly from falling down by gravity. The sequence to perform a calibration run is: 1) Powered on the motors. 2) Deploy the source to the defined destination. 3) Power off the motors. 4) Start data taking. In this way, noise introduced by motors was successfully avoided.

## 5. Position accuracy of the source

Many tests have been done before the MCS installation onto the AD. The assembly, shipping and installation procedures were finalized and practiced during these tests. The source position accuracy was also calibrated and surveyed in a simulated calibration environment named “Pit Test” (Fig. 10). During the “Pit Test”, a coordinate system relative to the support flange was well established and rulers and a Leica System 1200 Total Station [17] were used to ascertain the position accuracy of the MCS.

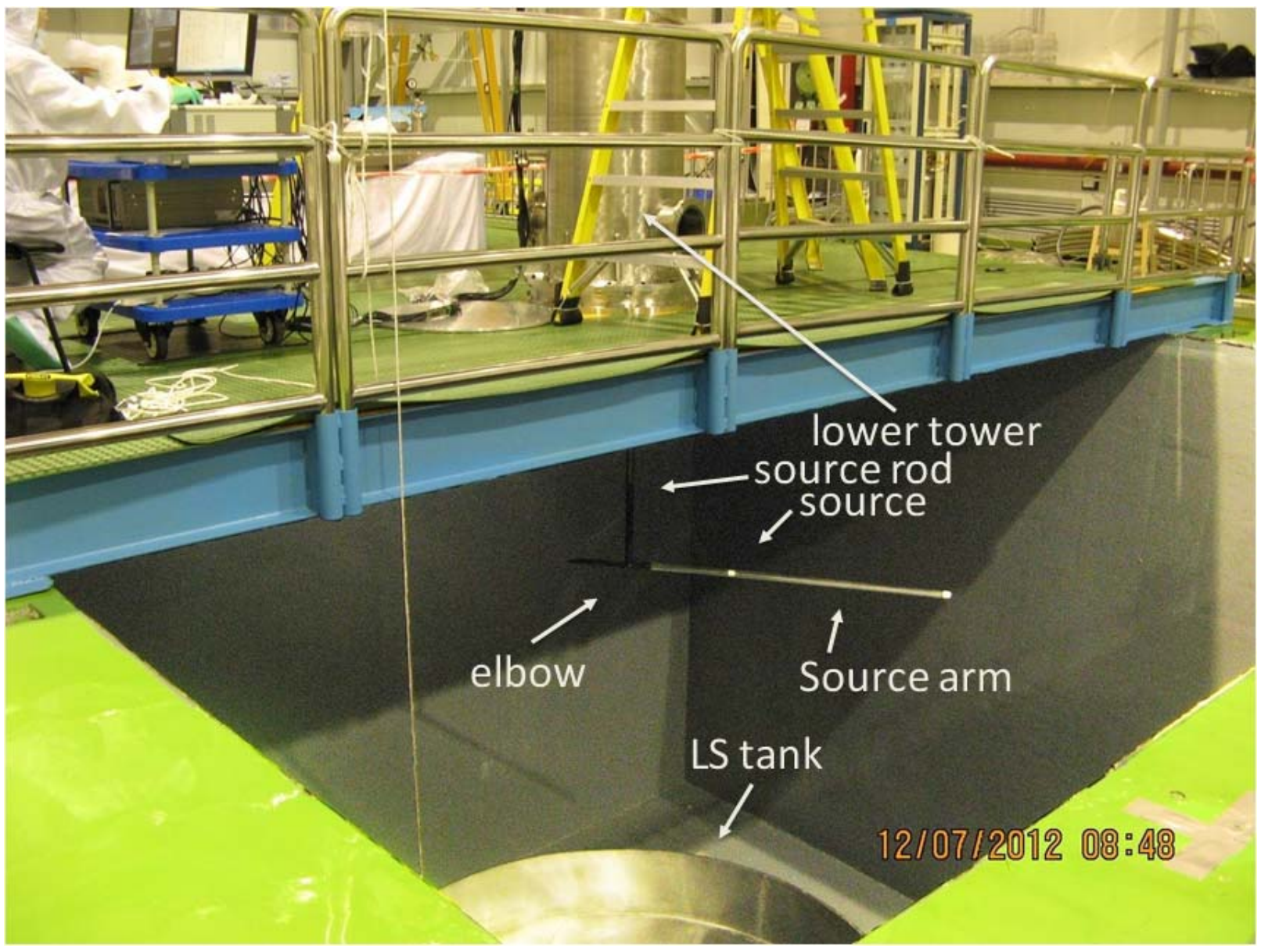

Fig. 10 Pit Test

For the R direction, the position accuracy was mainly affected by two factors: the motor control error and the off center of the source rod. The R position was feed back to computer with a rotational potential meter, which was calibrated with a ruler after the source was installed. The readout value varies within ±5 mm compared to the measurement of the ruler. After the MCS was installed on the AD, the central axis of the vertical source rod may deviate from the central axis of the AD. And the source rod may not be absolute straight and vertical. These lead to the off center of the source rod and it affected the R position accuracy directly. According to Leica’s measurement, the maximum off center value for the source rod is 24 mm. Therefore, the total position accuracy in R direction was calculated as $\Delta R=\mathrm{sqrt}(24^2+5^2)=25$ mm.

For Z axis, the position accuracy depends on three factors: the motor control error, the source arm deflection and the systematic error of AD. The motor control error was surveyed as 2 mm during the “Pit Test”. The acrylic source arm may deflect and affect the Z value. When the source arm was immersed into LS, the Z positions of the source were measured with Leica. The maximum bias in Z axis is 5 mm when source was deployed in different R values. The Z positions of IAV's various ports relative to the SSV top are mostly within 10 mm [5]. So, the total Z axis accuracy was calculated as $\Delta Z=\mathrm{sqrt}(2^2+5^2+10^2)=11.4\approx 12$ mm.

The Φ relative to the AD was also calibrated with Leica. The accuracy of Φ mainly depends on the installation error of MCS and was estimated at 0.5°.

Repeatability of the source was also measured during the Pit Test. For R and Z, a ruler was used as standard; for Φ, a scaled dial fixed on the rotation stage served as a standard. Table 1 summarized the position accuracy for different direction. One can see that the MCS can repeat source position to within 5 mm in R direction, 3 mm in Z axis and 0.2 ° in Φ direction. The source position accuracy is ΔR=25 mm, ΔZ =12 mm and ΔΦ=0.5°.

Table 1 Source position accuracy

| | R | Z | Φ |
|---|---|---|---|
| Range | 175—1350 mm | -1450—1200 mm | 0—350° |
| Position accuracy | 25 mm | 12 mm | 0.5 ° |
| Repeatability | 5 mm | 3 mm | 0.2 ° |

## 6. Operations

The calibration onsite was performed in AD1 which was deployed in the nearest experimental halls at over 1700 positions. A photograph of the MCS in operation is shown in Fig. 11. The calibration fully sampled the active region of the AD. Several detailed scans were taken, including around the circumference of the inner acrylic vessel at several heights, four scans in R-Z planes separated by 90 degrees, and special MCS configurations with the source arm vertical to position the source in the center of the detector. Throughout the calibration, several points were revisited to check for consistency and repeatability. The data acquisition time for each point varied from 5 to 120 minutes.

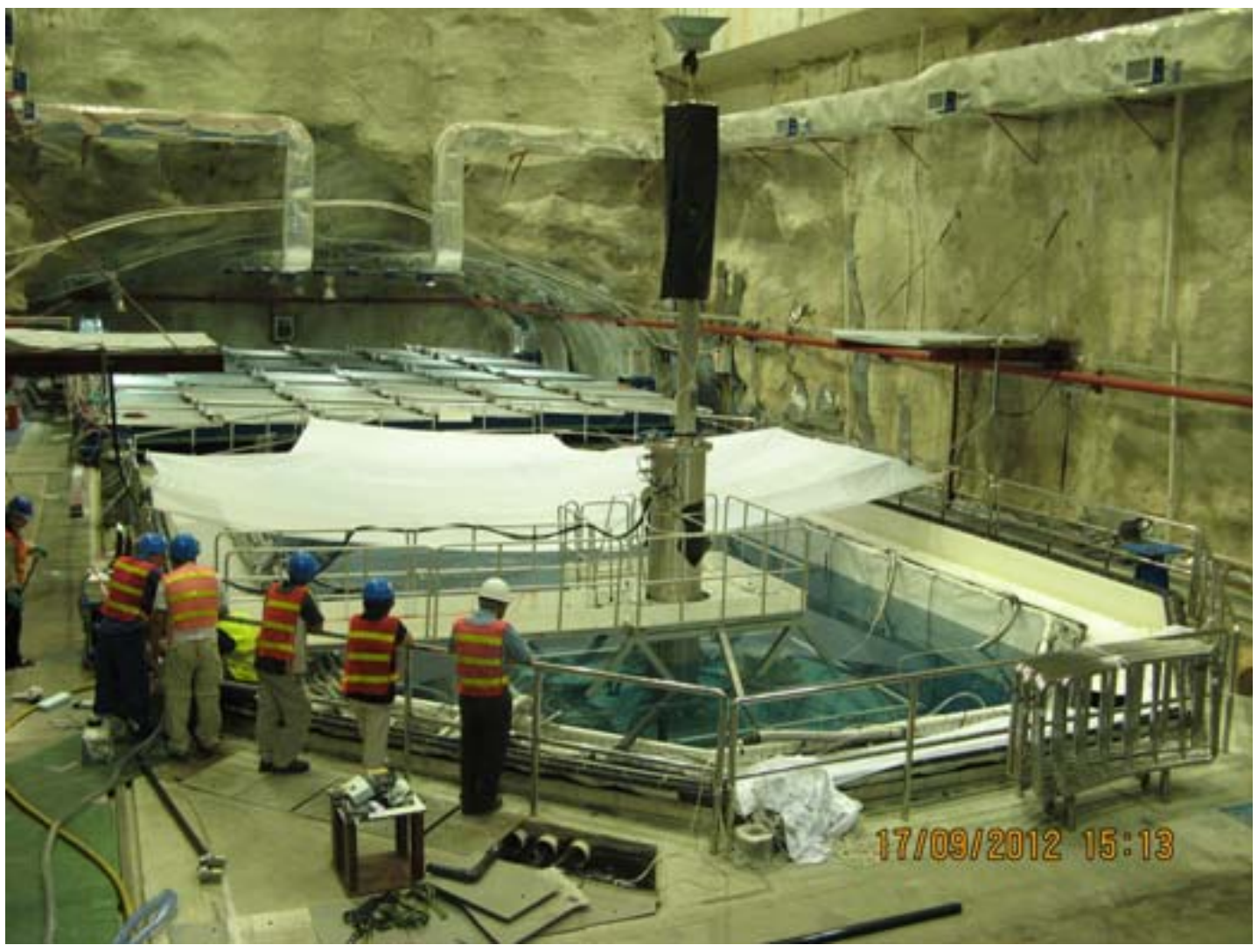

Fig. 11 Calibration on site

Fig. 12 shows the comparison of the detector spectrums before the MCS deployment and after that. The live time for both spectrums is 7000 second. After quenching, α decayed from $^{210}$Po would contribute to the peak around 0.5 MeV. The sum of a Gaussian and a linear function is utilized to fit the 0.5 MeV region to obtain the count rate of α from $^{210}$Po. The event rate of α is 9.75±0.04 Hz before MCS deployment and is 9.77±0.04 HZ after that. It is clear that no contaminations have been detected during the MCS calibration operations.

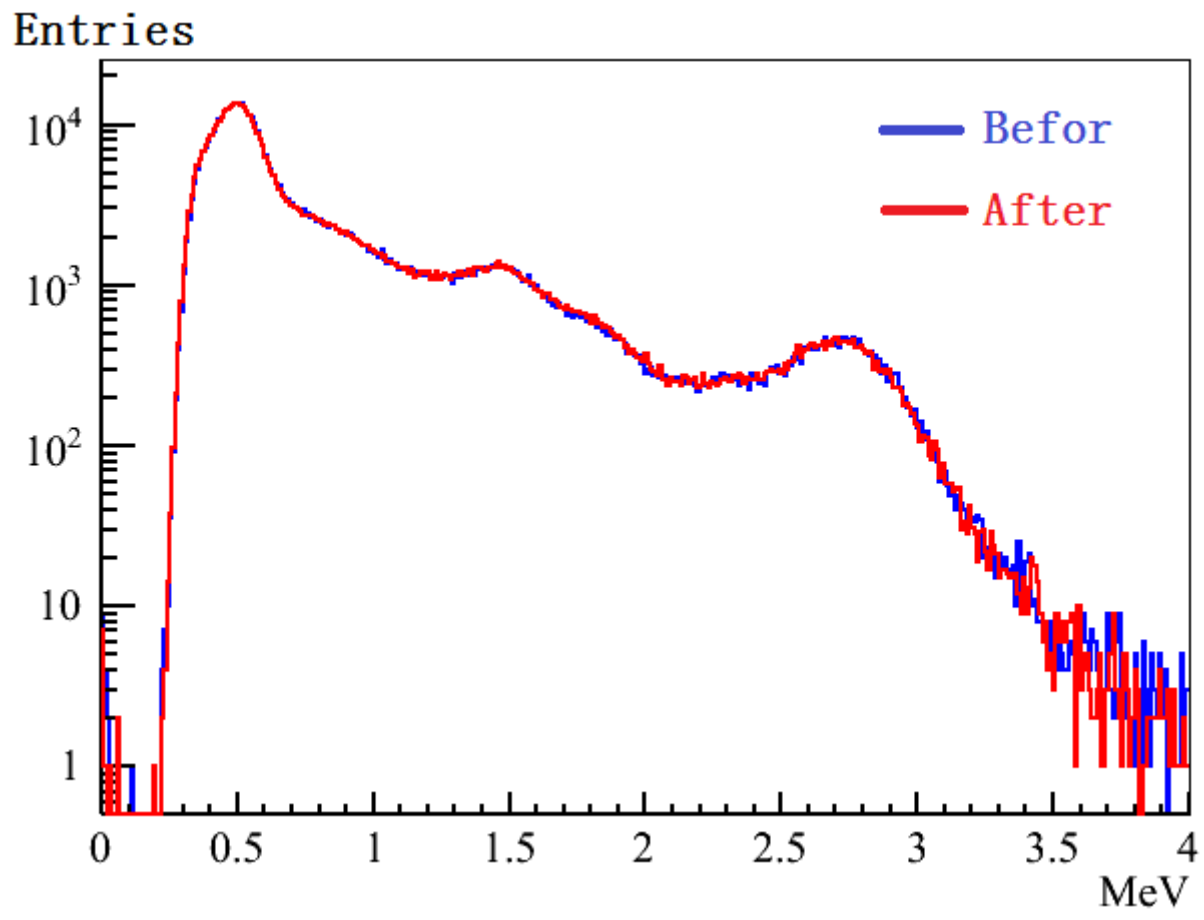

Fig. 12 Comparison of the detector spectrums within 7000 s before the MCS deployment and after that

## 7. Preliminary result

Since a lot of data has been acquired during the calibration and the data analysis is really complicated, it needs a lot of time to analyze all of it. Only preliminary result is presented in this paper.

The reproducibility of the MCS source position is shown in Fig. 13. Here the reconstructed position of the prompt signal from the $^{238}$Pu-$^{13}$C source is compared for MCS runs with the same nominal position; the shift in each Cartesian coordinate for each pair of runs is plotted as a function of radial distance from the center of the AD. The reconstructed position is determined by comparison to a set of Monte Carlo templates, using the method described in Ref. [9]. The MCS source position was changed many times between each pair of runs shown in Fig. 13, so from spread of the plotted points, it can be seen that the MCS position is reproducible to within 20 mm. This result is consistent with the measurement result during the Pit Test (Table 1).

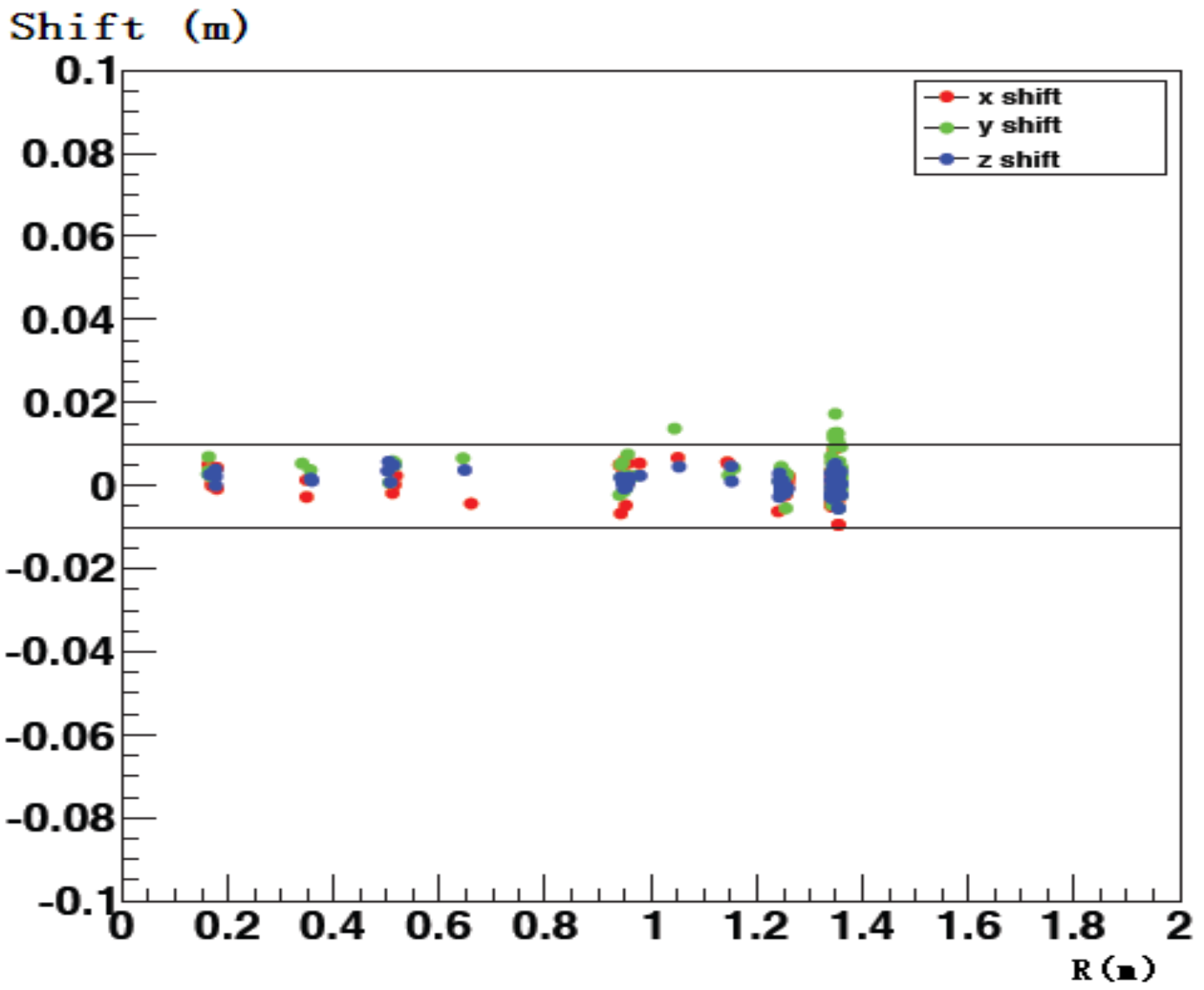

Fig. 13 Reproducibility of reconstructed MCS position: the difference between reconstructed positions in data for MCS runs with the same nominal position. The shifts are plotted as a function of radial distance from the AD center.

Fig. 14 shows an example of the calibration spectrums for prompt and delayed energies for two source positions. Events were selected by requiring delayed coincidence pairs similar to the IBD event selections. The black and blue plots are for a source position near the detector center (R=200 mm, Z=0 mm, Φ=240°). The red and purple plots are for source position near the edge of the active region (R=1350 mm, Z=0 mm, Φ=240°). The black and red plots show the prompt energies while the blue and purple plots show the delayed energies. Here, a simple scale factor is applied to convert raw photoelectrons into MeV with no position correction. Energies appear higher in the lower plots because more light is collected when the source is near the edge of the active region than at the center.

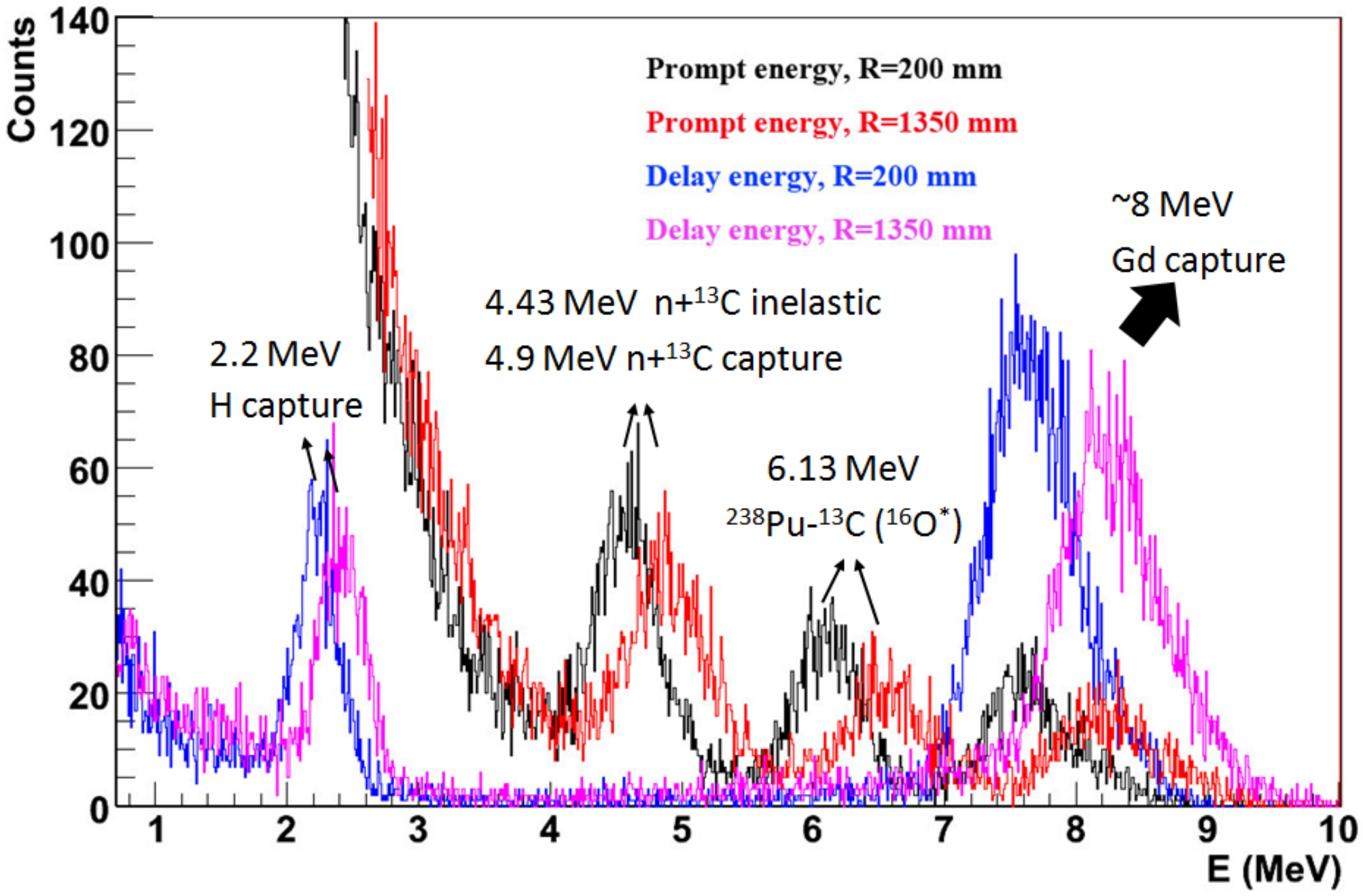

Fig. 14 An example of the calibration result

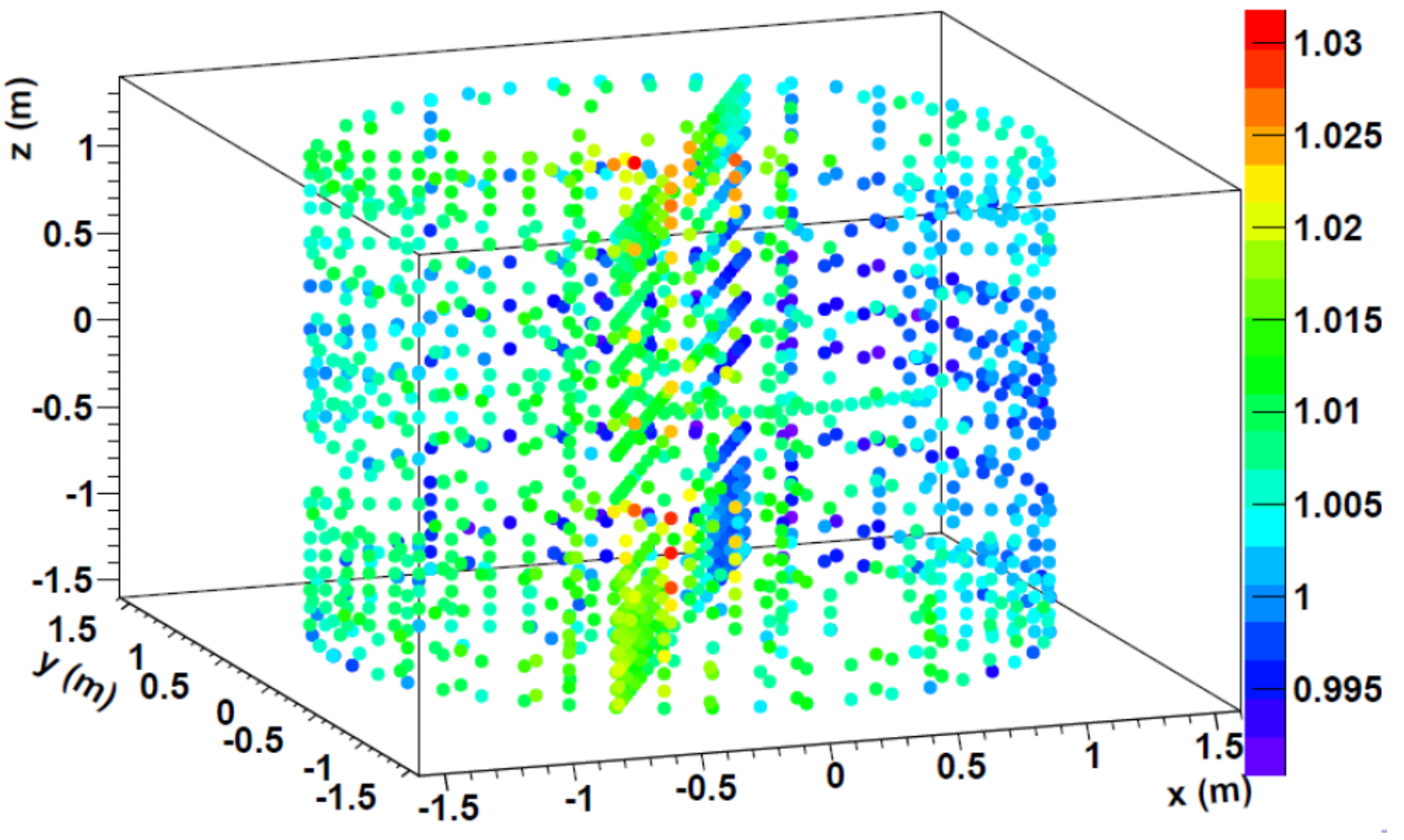

Fig. 15 8 MeV peak divided by the average 8MeV energy vs. source position

Fig. 15 shows a preliminary non-uniformity result of MCS. The energies from the Gd(n, γ) reaction in some positions are a little higher than 8 MeV and a little lower in some other positions. This owe to the variations of the AD response that has been explained in section 2 of this paper. This result didn’t consider the shadow effect of source rod and some other corrections.

## 8. Summary

It can be concluded that MCS position accuracy is a little poor than the design goals. But the position accuracy of $\Delta R=25$ mm, $\Delta Z=12$ mm and $\Delta\Phi=0.5^\circ$ is still quite good for such a complicated mechanism. The detail AD response study from the MCS calibration result will be published in the forthcoming paper. The system described here can also be applied to other system with similar characters where a full-volume sampling of a large cylindrical liquid volume is required. This paper is also an example of a successful developed robotic system with interlocked motions and remote control.

## 9. Acknowledgements

This work was supported by the National Natural Science Foundation under the grant No. 10890094 and the Ministry of Scinece and Technology of China under the grant No. 2013CB834306. The authors would like to thank Ralph Brown from BNL, Xiao Tang, Yuanguang Xia, Xiaoyan Ma, Linshu Wang, Jingyu Fu from IHEP, and many other technicians for their help during the MCS design, test and installation.